\makeatletter
\declare@file@substitution{revtex4-1.cls}{revtex4-2.cls}
\makeatother
\documentclass[twocolumn, times, tighten,twocolappendix]{aastex63}
\usepackage{amsmath,amssymb}
\usepackage[T1]{fontenc}
\usepackage{graphicx,multirow,hyperref,url,color,xspace}

\usepackage{rotating}
\usepackage{subfigure}

\newcommand{\msun}{{M}_{\sun}}

\newcommand{\ledd}{L_{{\rm Edd}}}

\newcommand{\xte}{{\textit{RXTE}}\xspace}

\newcommand{\suzaku}{{\textit{Suzaku}}\xspace}
\newcommand {\magenta}[1] {\textcolor{black}{#1}}
\newcommand{\nustar}{{\textit{NuSTAR}}\xspace}
\newcommand{\nicer}{NICER\xspace}

\newbox\grsign \setbox\grsign=\hbox{$>$} \newdimen\grdimen \grdimen=\ht\grsign
\newbox\simpropbox
\setbox\simpropbox=\hbox{\raise.5ex\hbox{$\propto$}\llap
 {\lower.5ex\hbox{$\sim$}}}\ht2=\grdimen\dp2=0pt
\def\simprop{\mathrel{\copy\simpropbox}}

\begin{document}

\title{Black hole spin measurements in LMC X-1 and Cyg X-1 are highly model-dependent}

\author[0000-0002-0333-2452]{Andrzej A. Zdziarski}
\affiliation{Nicolaus Copernicus Astronomical Center, Polish Academy of Sciences, Bartycka 18, PL-00-716 Warszawa, Poland; \href{mailto:aaz@camk.edu.pl}{aaz@camk.edu.pl}}

\author[0000-0002-6051-6928]{Srimanta Banerjee}
\affiliation{Inter-University Center for Astronomy and Astrophysics, Pune 411007, India}

\author[0000-0003-3499-9273]{Swadesh Chand}
\affiliation{Inter-University Center for Astronomy and Astrophysics, Pune 411007, India}

\author[0000-0003-1589-2075]{Gulab Dewangan}
\affiliation{Inter-University Center for Astronomy and Astrophysics, Pune 411007, India}

\author[0000-0002-7609-2779]{Ranjeev Misra}
\affiliation{Inter-University Center for Astronomy and Astrophysics, Pune 411007, India}

\author[0000-0001-7606-5925]{Micha{\l} Szanecki}
\affiliation{Faculty of Physics and Applied Informatics, {\L}{\'o}d{\'z} University, Pomorska 149/153, PL-90-236 {\L}{\'o}d{\'z}, Poland}

\author[0000-0002-8541-8849]{Andrzej Nied{\'z}wiecki}
\affiliation{Faculty of Physics and Applied Informatics, {\L}{\'o}d{\'z} University, Pomorska 149/153, PL-90-236 {\L}{\'o}d{\'z}, Poland}

\begin{abstract}
The black-hole spin parameter, $a_*$, was measured to be close to its maximum value of 1 in many accreting X-ray binaries. In particular, $a_*\gtrsim 0.9$ was found in a number of studies of LMC X-1. These measurements were claimed to take into account both statistical and systematic uncertainties. We perform new measurements using a recent simultaneous observation by NICER and NuSTAR, providing a data set of high quality. We use the disk continuum method together with improved models for coronal Comptonization. With the standard relativistic disk model and optically thin Comptonization, we obtain values of $a_*$ similar to those obtained before. We then consider modifications to the standard model. Using a color correction of 2, we find $a_*\approx 0.64$--0.84. We then consider disks with dissipation in surface layers. To account for that, we assume the standard disk is covered by a warm and optically thick Comptonizing layer. Our model with the lowest $\chi^2$ yields then $a_*\approx 0.40^{+0.41}_{-0.32}$. In order to test the presence of such effects in other sources, we also study an X-ray observation of Cyg X-1 by Suzaku in the soft state. We confirm the previous findings of $a_*>0.99$ using the standard model, but then we find a weakly constrained $a_*\approx 0.82^{+0.16}_{-0.74}$ when including an optically thick Comptonizing layer. We conclude that determinations of the spin using the continuum method can be highly sensitive to the assumptions about the disk structure.
\end{abstract}

\section{Introduction} \label{intro}

One of the three parameters of a black hole (BH) is its spin. In the case of BH X-ray binaries (hereafter BHXRBs), its measurements are based on two methods. One, done mostly in the hard spectral state (see, e.g., \citealt{DGK07} for definition of spectral states), uses fits of reflection spectra, including the Fe K fluorescent line (see \citealt{Bambi21} for a review). This method relies on the strong dependence of the extension of the red tail of relativistically broadened lines on the emitting radii, as well as on the dependence of the radius of the innermost stable orbit (ISCO) on the BH spin. The method appears to be controversial because the measurements depend on the shape of the assumed underlying continuum as well as they assume that the accretion disk extends to either the ISCO or close to it, which is uncertain in the hard spectral state (e.g., \citealt{Basak16, Mahmoud19, Kawamura22, Kawamura23}). The other method relies on the shape of the disk continuum in the soft spectral state (see \citealt{McClintock14} for a review). It assumes that the disk extends to the ISCO, which, for the soft state, is well founded observationally  (e.g., \citealt{DGK07} and references therein). Unlike the previous method, it requires the knowledge of the distance to the source and the BH mass, which is a feature of this method. The method can then be used for estimating the distance and mass for accurate enough data. Then, the observed spectra usually include some high-energy tails, appearing to be due to Comptonization of the disk photons by (at least mildly) relativistic electrons. The results published with the latter method usually include that effect using phenomenological models, either {\tt simpl} \citep{Steiner09} or a power law. As we find in this work, the effect of Comptonization on the spin determination can be quite major, which is another feature of this method.

With both methods, the spins of many BHXRBs were found to be high, with the dimensionless spin parameter of $a_*\gtrsim 0.9$, see, e.g., the compilation in \citet{Fishbach22}. The high values of the spin claimed in many BHXRBs differ strongly from those inferred for binary BHs from the majority (70--90\%) of the merger events observed in gravitational waves \citep{Abbott23}. In those measurements, the effective spin parameter, which is an average of the individual spins weighted by both the masses and the misalignment angles, \magenta{has the mean value of $\approx 0.06$}. This prompted \citet{Fishbach22} to propose that these two groups of BH binaries form populations with different formation scenarios and/or origins. On the other hand, \citet{Belczynski23} used the metallicity-dependent population synthesis code of \citet{Belczynski08, Belczynski10} and was able to explain both low-mass BHs formed at high metallicities and high-mass BHs formed at low metallicities, and the distribution of the spins measured via BH mergers. 

If the spins of the BHXRBs are high on average, then we need to understand its origin. Accretion itself is inefficient in spinning up BHs in either low or high-mass XRBs, see, e.g., \citet{Belczynski20}. On the other hand, theoretical predictions for the natal spins at BH masses of $\sim\!\! 10\msun$ vary from $a_*\ll 1$ for models with super-efficient angular momentum transfer \citep{Fuller19} through $a_*\sim 0.1$ for standard MESA stellar models with efficient transport \citep{Spruit99, Spruit02} to $a_*\sim 0.8$--0.9 for Geneva stellar models with moderate angular momentum transfer \citep{Ekstrom12}; see \citet{Belczynski20} for a review. As argued in that paper, the low average spins in binary BHs favor relatively low natal spins of BHs in the majority in the accreting systems (though high spins in some cases remain possible).  

\begin{table*}\centering
\caption{The log of the studied observations with the \nicer and \nustar} 
%\vskip -0.4cm                               
\begin{tabular}{cccccc}
\hline
Detector & Obs. ID & Start time [MJD] & End time [MJD] & Exposure [s] & Phase\\
\hline
NICER & 5100070106 & 59876.36971 & 59876.37602 & 545 & 0.055--0.056\\
 & -- & 59876.69251 & 59876.70388 & 982 & 0.137--0.140\\
NuSTAR A & 90801324002 &59876.16399 & 59876.60497 & 19280 & 0.002--0.115\\
NuSTAR B & -- & 59876.16399 & 59876.60497 & 19130 & 0.002--0.115\\
\hline
\end{tabular}
\label{log}
\tablecomments{The start/end times for the NICER correspond to the two segments of the data used for the fits and the power spectrum. The orbital phases are according to the ephemeris of \citet{Orosz09}, with the null phase corresponding to the BH behind the supergiant.}
\end{table*}

\begin{figure}
\centerline{\includegraphics[width=\columnwidth]{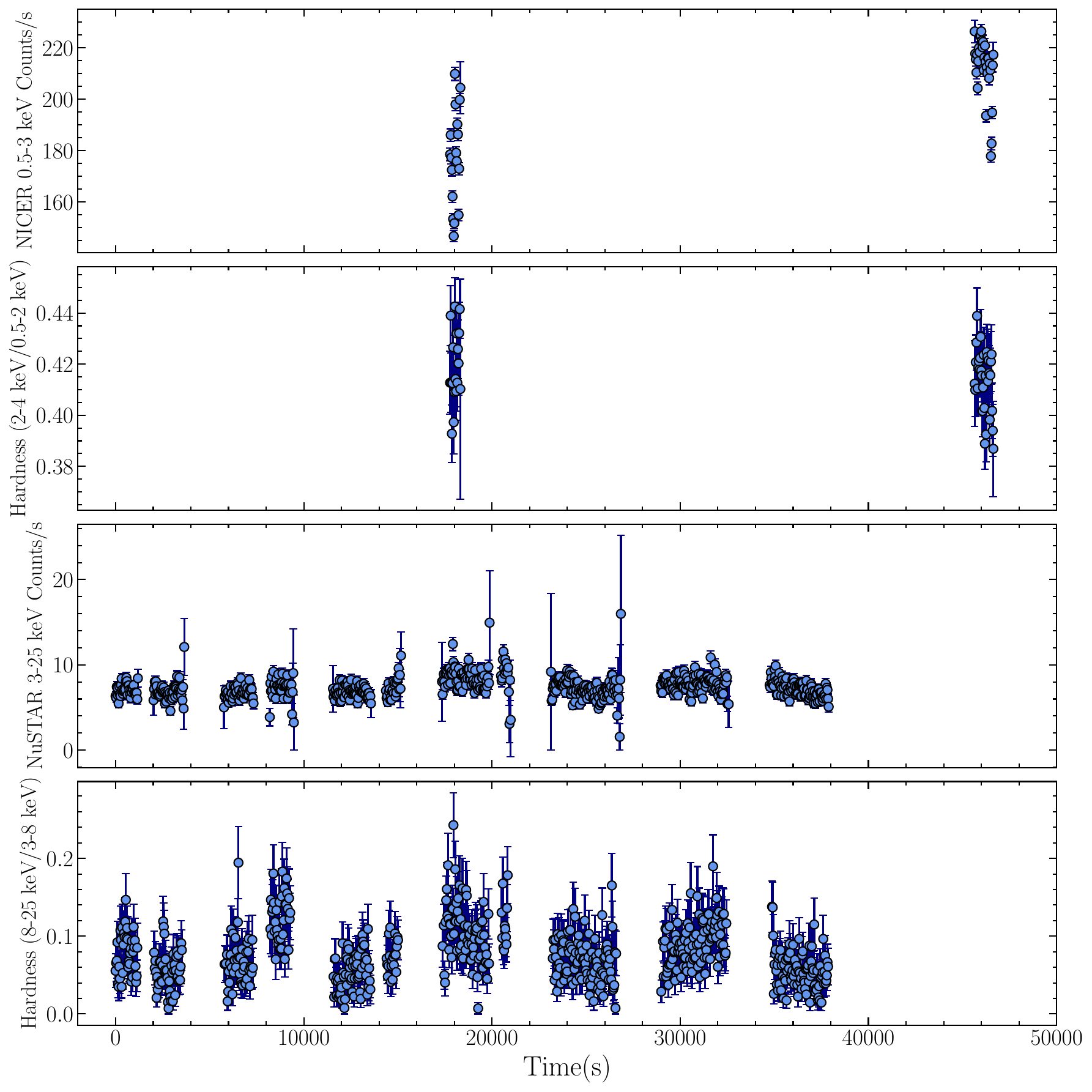}}
  \caption{The count rates and hardness ratios vs.\ time for the LMC X-1 data used here. The top two panels show the 0.5--3.0\,keV count-rate and the (2--4\,keV)/(0.5--2.0\,keV) count rate ratio. The bottom two panels show the 3--25\,keV count-rate and the (8--25\,keV)/(3--8\,keV) count rate ratio.
}\label{lc}
\end{figure}

\begin{figure}
\centerline{\includegraphics[width=8cm]{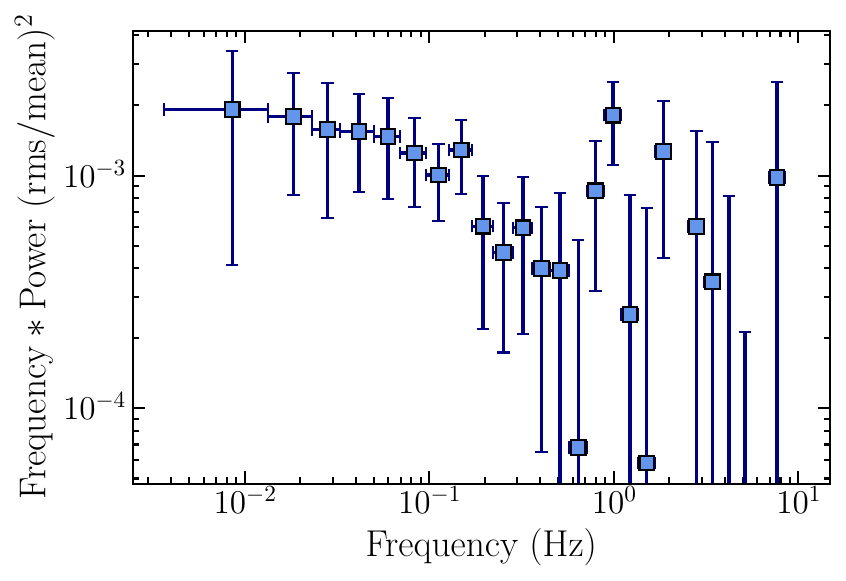}}
\centerline{\includegraphics[width=8cm]{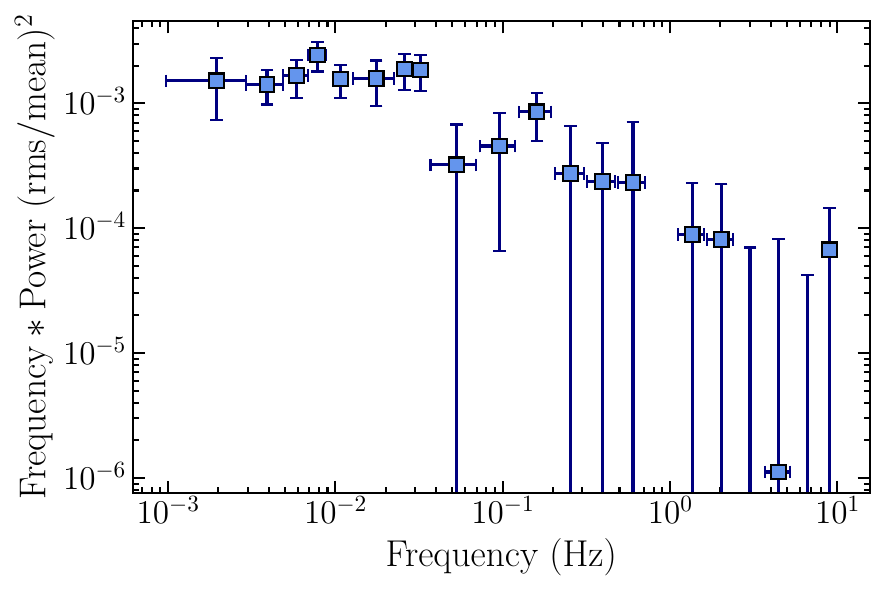}}
  \caption{Top: the power spectra from NICER, 0.34--9.0 keV. Bottom: the co-spectra from \nustar, 3.0--30 keV. 
}\label{power}
\end{figure}

In this work, we concentrate on the case of the high-mass BHXRB LMC X-1, which has always been observed to be in the soft spectral state (e.g., \citealt{GD04}). Its distance and BH mass are well known, see Section \ref{results}. We consider the issue of the reliability and model dependence of its spin measurements with the continuum method. We use the simultaneous X-ray observations by \nicer and \nustar in 2022 \citep{Podgorny23}, see Table \ref{log}. The combination of \nicer and \nustar provides a highly accurate data set, allowing us to determine the spectral parameters very accurately. The BH spin in LMC X-1 was first studied using the continuum method including the general-relativistic (GR) effects by \citet{Gierlinski01}, whose results were, however, not conclusive. It was then measured by \citet{Gou09} with the same method using a large set of \xte PCA spectra, for which they found $a_*\approx 0.92^{+0.05}_{-0.07}$. Similarly high values were found by \citet{Steiner12a, Mudambi20, Jana21, Bhuvana22}. In \citet{Gou09}, reflection of the Comptonization emission from the disk was neglected, while it was included in \citet{Steiner12a}. In \citet{Mudambi20} and \citet{Bhuvana22}, reflection was included in a simplified way by adding a broad line to the model. In \citet{Jana21}, relativistic reflection was included in some models, but it resulted only in upper limits on the reflection fraction. In the present work, we fit our data with usual models of GR disks and coronal Comptonization, but we also consider the possible presence of a warm skin on top of the disk.

In addition, we consider Cyg X-1 as a comparison source, see Appendix \ref{cygx1}. The source was discovered in 1964 \citep{Bowyer65}, and it is probably the best studied BHXRB to date. Similarly to LMC X-1, it has a high-mass donor and accurate determinations of its binary parameters and the distance \citep{Miller-Jones21}, see Appendix \ref{cygx1}. Also, its spin has been claimed to be extremely high \citep{Miller-Jones21}. We fit the data acquired by \suzaku in one of its softest states ever observed \citep{Kawano17}. We apply to those data the same approach as for LMC X-1.

\section{The data} \label{data}

Our data set consists of approximately simultaneous broad-band X-ray observations of LMC X-1 in the soft state by \nicer and \nustar in 2022 \citep{Podgorny23}. The \nicer observation contemporaneous to that of \nustar was in three short segments, and the last two were after the end of the \nustar observation. Also, the spectral hardness decreased with time, which effect was most pronounced for the third segment. Consequently, we have found that simultaneous fitting of the \nustar data with the full \nicer data required a significant correction for the overall hardness of either of the instruments (which we correct for using a power-law model, see Section \ref{model} below), $\Delta\Gamma\approx 0.1$. However, using only the first segment would strongly reduce the exposure, by a factor of 5. On the other hand, using the first two segments of the \nicer observation reduced their exposure by 29\% only, and gave a very good agreement with the \nustar spectra in the overlapping region, with only a tiny slope correction required, $\Delta\Gamma< 0.01$. Thus, we use only the first two segments of the \nicer observation. The log of the used data is given in Table \ref{log}, and the count rates and hardness ratios are shown in Figure \ref{lc}. We see that while the \nicer count rate moderately increased during the second data segment, the hardness ratios remained similar. In case of the \nustar data, both the count rates and hardness ratio remained rather stable.

The source flux shows low variability on short time scales. We have calculated the power spectra in the 0.34--9.0 (\nicer) and 3--30\,keV (\nustar) energy ranges with the white noise subtracted, see Figure \ref{power}. The range of the former is dominated by the thermal disk component, see Figure \ref{eeuf_ratio} below.  The \nicer power spectrum is created with a time bin size of 0.05\,s (giving the Nyquist frequency of 10\,Hz). We then generate intervals of $8192\times 0.05\approx 410$\,s, and generate the power density spectrum for each of them. The spectra of different intervals are averaged and the resulting spectrum are logarithmically rebinned with a factor 1.05 in frequency to improve statistics. The latter is the co-spectrum calculated using the {\sc hendrics} package \citep{Bachetti15, Lazar21}. We see that the fractional variability squared per unit $\ln \nu$ is low, $\sim\! 2\times 10^{-3}$. We have computed the rms in the 0.01--1\,Hz range, and found it to be $5.3\pm 0.1\%$ and $10.3\pm 0.1\%$, respectively. Thus, both the disk blackbody and the high-energy tail are quite stable.

The \nicer and \nustar data used for the spectral analysis are in the 0.34--9.0 and 3--41\,keV ranges, respectively. Above the \nustar upper limit, the spectrum is background-dominated. During the standard data reduction of NICER, the script {\tt nicerl3-spect} added 1.5\% systematic error to each channel. We have not added any systematic errors to the \nustar data. Both of the spectral data have been optimally binned \citep{Kaastra16} requiring a minimum of 20 counts per channel.

\section{The models} \label{model}

\begin{table*}
\caption{The results of spectral fitting
}
   \centering\begin{tabular}{lcccccc}
\hline
Component & Parameter & Model 1 & Model 2 & Model 3 & Model 4 & Model 5\\
\hline
LMC absorption & $N_{\rm H,LMC}$ $[10^{22}]$\,cm$^{-2}$ & $0.97^{+0.01}_{-0.01}$ & $0.99^{+0.02}_{-0.01}$ & $0.97^{+0.01}_{-0.01}$ & $0.98^{+0.03}_{-0.02}$ & $0.97^{+0.01}_{-0.01}$\\
\hline
Disk & $a_*$ & $0.90^{+0.02}_{-0.01}$ & $0.94^{+0.02}_{-0.04}$ & $0.77^{+0.07}_{-0.13}$ & $0.40^{+0.41}_{-0.32}$ & $0.72^{+0.02}_{-0.33}$\\
& $\dot M_{\rm disk}\, [10^{18}{\rm g\,s}^{-1}]$ & $1.8^{+0.1}_{-0.1}$ & $1.3^{+0.5}_{-0.5}$ & $1.8^{+0.8}_{-0.4}$ & $1.03^{+1.10}_{-0.01}$ & $0.96^{+0.31}_{-0.04}$\\
\hline
Coronal  & $y$ & $0.20^{+0.03}_{-0.02}$ & $0.09^{+0.11}_{-0.03}$  & $0.11^{+0.07}_{-0.04}$ & $0.075^{+0.127}_{-0.002}$ & $0.072^{+0.016}_{-0.002}$ \\
 Comptonization & $kT_{\rm e}$ [keV] & $11^{+56}_{-4}$ &  $50^{+60}_{-14}$ &  $50^{+27}_{-9}$ &  $52^{+10}_{-7}$ &  $61^{+11}_{-10}$\\
& $\tau_{\rm T}$ & 2.3 & 0.23 & 0.28 & 0.18 & 0.15\\
&$f_{\rm cov}$ & $0.1^{+0.3}_{-0.1}$ &  $0.3^{+0.5}_{-0.2}$ &  $0.4^{+0.6}_{-0.3}$ &  $1.0_{-0.9}$&  $1.0_{-0.9}$\\
\hline 
Disk skin & $\tau_{\rm T}$ & -- & -- & -- & $24^{+2}_{-7}$ & $18.1^{+0.3}_{-17.1}$ \\
Comptonization & $kT_{\rm e}$ [keV] & -- & -- & -- & $0.84^{+0.10}_{-0.02}$ & $0.94^{+0.04}_{-0.04}$ \\
\hline
Fe K line/ &$E_{\rm line}$ [keV] & $6.3^{+0.6}_{-0.3}$ & -- & $6.2^{+0.6}_{-0.2}$ & $6.1^{+0.6}_{-0.1}$ & $6.95^{+0.02}_{-0.15}$\\
reflection &$\sigma_{\rm line}$ [keV]  & $1.1^{+0.4}_{-0.4}$ & -- & $1.1^{+0.7}_{-0.4}$ & $1.3^{+0.2}_{-0.4}$ & -- \\
& $N_{\rm line}\,[10^{-4}\,{\rm cm}^{-2}{\rm s}^{-1}]$ & $3^{+2}_{-2}$ & -- & $3^{+17}_{-2}$ & $5^{+2}_{-2}$ & $1.3^{+1.0}_{-0.5}$ \\
&$\log_{10}\xi$ & -- & $3.5^{+0.5}_{-0.6}$ &  -- & --\\
  & $kT_{\rm diskbb}$ [keV] & -- &  $0.6^{+0.1}_{-0.4}$ & -- & --\\
& $N_{\rm refl}\,[10^{-2}\,{\rm cm}^{-2}{\rm s}^{-1}]$ & -- & $0.03^{+0.02}_{-0.01}$ & -- & --\\
\hline
Cross-calibration & $\Delta\Gamma_{\rm NICER}$ & $0.01^{+0.03}_{-0.03}$ & $0^{+0.03}_{-0.04}$ & $0.01^{+0.03}_{-0.02}$ & $0.01^{+0.05}_{-0.04}$ & $0.01^{+0.05}_{-0.05}$\\
& $K_{\rm NICER}$ & $1.00^{+0.05}_{-0.02}$ & $1.00^{+0.05}_{-0.06}$& $1.01^{+0.05}_{-0.06}$& $1.01^{+0.07}_{-0.06}$ & $1.01^{+0.07}_{-0.07}$ \\
& $\Delta\Gamma_{\rm NuSTAR B}$ & $-0.02^{+0.03}_{-0.03}$ & $-0.02^{+0.03}_{-0.03}$ & $-0.02^{+0.03}_{-0.03}$ & $-0.03^{+0.03}_{-0.03}$ & $-0.03^{+0.03}_{-0.02}$\\
&$K_{\rm NuSTAR B}$ & $0.95^{+0.05}_{-0.04}$ & $0.95^{+0.04}_{-0.04}$ & $0.95^{+0.05}_{-0.04}$ & $0.95^{+0.05}_{-0.04}$ & $0.95^{+0.05}_{-0.04}$\\
\hline
& $\chi_\nu^2$  & 245/265 & 250/265 & 246/265 & 242/263 & 246/264\\
\hline
\end{tabular}
\tablecomments{Model 1 = {\tt plabs*tbabs*tbvarabs*comppsc(kerrbb2+gaussian)}; Model 2 = {\tt plabs*tbabs*tbvarabs*comppsc(kerrbb2+reflkerr)}; Model 3 = {\tt plabs*tbabs*tbvarabs*comppsc(kerrbb+gaussian)} with $\kappa=2$; Model 4 = {\tt plabs*tbabs*tbvarabs*comppsc(thcomp(kerrbb2)+gaussian)}; Model 5 = {\tt plabs*tbabs*tbvarabs*comppsc(thcomp(kerrbb2)+reflkerrline)}. In coronal Comptonization, $y\equiv 4(k T_{\rm e}/m_{\rm e}c^2)\tau_{\rm T}$, the $\tau_{\rm T}$ is given at the best fit (not a free parameter). $\xi\equiv 4{\pi}F_{\rm{irr}}/n$ in Model 2 is the ionization parameter of the reflector, where $F_{\rm{irr}}$ is the irradiating flux in the 0.1--1000\,keV range. The energies of the Gaussian feature and the {\tt reflkerrline} are limited to $6.0\leq E_{\rm line}/{\rm keV}\leq 7.0$ and $6.40\leq E_{\rm line}/{\rm keV}\leq 6.97$, respectively. The inclination of $i=36\fdg 4$ and $M_{\rm BH}=10.9\msun$ are assumed.}
\label{fits}
\end{table*}

The distance to the center of the LMC is $49.6\pm 0.6$\,kpc \citep{Pietrzynski19}, and we assume here the distance to LMC X-1 of 50 kpc.The BH and donor masses of LMC X-1 were measured as $M_{\rm BH}=10.9\pm 1.4$ and $M_*=31.8\pm 3.5\msun$, respectively, and the orbital period and inclination as $P=3.90917 \pm 0.00005$\,day and $i=36\fdg 4 \pm 1\fdg 9$, respectively \citep{Orosz09}. 

In our spectral analysis, we use the X-ray fitting package {\sc{xspec}} \citep{Arnaud96}. The fit uncertainties are for 90\% confidence, $\Delta\chi^2 \approx 2.71$ \citep{Lampton76}. Residual differences between the calibration of the different detectors are accounted for by the model {\tt plabs}, which multiplies the model spectra by $K E^{-\Delta\Gamma}$. We set $K$ and $\Delta\Gamma$ fixed at 1 and 0, respectively, for the \nustar A module. We allow $\Delta\Gamma\neq 0$ for \nustar B to allow for residual differences with respect to A, which differences are, however, rather small \citep{Madsen22}. Since we use convolution models, we use the {\sc xspec} command {\tt 'energies 0.001 2000 2000 log'} to set the energy grid.

The Galactic column density toward the source is $N_{\rm H,Gal}\approx 5.9\times 10^{20}$\,cm$^{-2}$ \citep{HI4PI}, while the column through the LMC toward LMC X-1 is much larger, $N_{\rm H,LMC}\approx (1.0$--$1.3)\times 10^{22}$\,cm$^{-2}$ \citep{Hanke10}, which includes the contribution of the donor stellar wind. For the Galactic absorption, we assume the abundances of \citealt{Wilms00}, as given in their model {\tt tbabs}. The elemental abundances in the LMC are lower than those in the Galaxy. We initially adopted those listed in table 2 of \citet{Hanke10}, and, in the case of the elements not listed there, we assumed a half of the Galactic abundances, using the model {\tt tbvarabs} \citep{Wilms00}. However, we found then strong residuals at energies $\lesssim$0.5\,keV in all of our models. We have then fitted the O and Fe fractional abundances and found the best fits at 0.80 and 0.48, respectively, which significantly reduced (but not removed) the residuals. We have thus kept them fixed at these values hereafter.

For modelling of optically-thick spectra of accretion disks around a Kerr BH, we use either the {\tt kerrbb} model \citep{Li05} or {\tt kerrbb2} \citep{McClintock06}. The former is a GR model with the local emission modelled as blackbody with a color correction, $\kappa$ \citep{ST95, Davis19}, as a parameter. The latter uses the GR treatment of {\tt kerrbb} but it calculates local disk spectra based on the radiative transfer calculations of \citet{Davis05,Davis06} as given in their model, {\tt bhspec}, available for the two values of the viscosity parameter, $\alpha=0.01$ and 0.1.

The spectra observed from LMC X-1 show significant high-energy tails beyond the disk components. They appear to be due to Comptonization by electrons in a corona covering the disk (partially or fully). We use a convolution version, {\tt comppsc}, of the {\tt compps} iterative-scattering code \citep{PS96}, assuming a slab geometry. This is a highly accurate Comptonization code, calculating angle-dependent spectra from plasmas at arbitrary optical depths (though the used method causes calculations at high Thomson optical depths, $\tau_{\rm T}\gtrsim 5$, to be very slow). The seed photons for Comptonization are those from the accretion disk, calculated as described above. When high-energy tails are observed at high enough energies from BHXRBs, it has been found that the distribution of Comptonizing electrons is hybrid, i.e., Maxwellian with a high energy tail \citep{G99, ZGP01, McConnell02, GD03, GZ03}. The {\tt compps} model allows for such tails. However, the spectra studied in this work can be significantly measured only up to $\sim$30--40\,keV, and we have found that thermal Comptonization is sufficient to model them. The parameters of the model are the electron temperature, $kT_{\rm e}$, the Compton parameter, defined as $y\equiv (kT_{\rm e}/m_{\rm e}c^2)\tau_{\rm T}$, and the corona covering fraction, $f_{\rm cov}$. 

In addition, we consider a possibility that the surface of the disk is partially supported by magnetic pressure \citep{Begelman07, Salvesen16a, Begelman17, Dexter19, Mishra20}. To account for that, we \magenta{increase the color correction in some cases to $\kappa=2$ \citep{Salvesen21}. Furthermore, we consider the possibility that a disk surface layer is dissipative, which we take into account by assuming} in some models that the disk is fully covered by a warm scattering plasma with $kT_{\rm e}\sim 1$\,keV and the Thomson optical depth of $\tau_{\rm T}\gg 1$. \magenta{We model it} using the {\tt thcomp} convolution model \citep{Z20_thcomp}. The parameters of the model are the electron temperature, $kT_{\rm e}$, the Thomson optical depth, $\tau_{\rm T}$, and the covering fraction, which we fix at unity.

Our Comptonization models, {\tt comppsc} and {\tt thcomp}, are convolved with either {\tt kerrbb} or {\tt kerrbb2}, which give the observed emission, following the relativistic effects of photon propagation. Strictly, we should apply the Comptonization in the local frames, and then apply the GR correction; however, such models are currently not available. We note that the two effects, Comptonization and GR effects on the photon propagation, would be strictly commutative if the photon paths to the observer from the surface of both the disk and the corona were identical. In the case of {\tt thcomp}, the optical depth of the plasma is large and the angular distribution of the escaping photons is almost the same as of the blackbody-emitting disk. Then, the photon paths would not be affected by Comptonization for the scale height of the skin being at most moderate, $H/R\lesssim 0.1$. In the case of {\tt comppsc}, the plasma has $\tau_{\rm T}\lesssim 2$, and thus the angular distribution of the escaping photons is moderately different from the optically thick case. However, only a fraction $f_{\rm cov}[1-\exp(-\tau_{\rm T})]\sim 0.1$ of the disk/skin photons is Comptonized. Then, a part of the scattered photons goes into the high-energy tail, which is mostly a featureless power law, where the accuracy of the GR corrections is not critical. Thus, the Comptonization and the GR corrections are in our case approximately commutative, i.e., their order can be reversed with little loss of accuracy.

The spectra also show broad Fe K$\alpha$ lines, indicative of relativistic reflection. We model it either phenomenologically as a broad Gaussian feature (following \citealt{Mudambi20, Jana21, Bhuvana22}) or using the {\tt reflkerr} relativistic reflection model \citep{Niedzwiecki19}. In both cases, the reflection feature undergo Comptonization in the corona in the same way as the direct disk emission.

The {\tt reflkerr} model uses the incident spectra from Comptonization calculated using {\tt compps} and allows the user to choose the slab geometry, consistently with our treatment of the disk Comptonization. The model also allows the disk irradiation profile to be either a power-law or the same as the disk emissivity \citep{Page74}. The latter corresponds to the case of a corona with the internal dissipation given by a constant fraction of the disk dissipation and a low scale height. On the other hand, power-law profiles, commonly used in other works, appear not to be consistent with the assumption of the zero stress (and zero dissipation) at the ISCO, since they have the maxima at the ISCO. As an alternative, we use a model of an extended (and rotating with the disk) corona, with the height as a free parameter, in which the disk irradiation is calculated self-consistently \citep{Szanecki20, Klepczarek23}, {\tt reflkerrvnth}. 

Since these models use tabulated rest-frame reflection profiles of {\tt xillverCp} \citep{GK10, Garcia18}, in which the ionizing flux is low ($\propto$ the reflector density, assumed to be $n=10^{15}$\,cm$^{-3}$), the absorbed and re-emitted photons appear mostly below a few hundred eV, i.e., below the observed energy range \citep{Garcia16, ZDM20}. Also, {\tt xillverCp} neglects effects of the disk temperature, which is high in our case. Thus, we effectively neglect the re-emission of the absorbed part of the irradiating flux. However, since the albedo for high disk temperatures is close to unity (due to collisional ionization) and the reflection components constitute less than a few \% of the total flux (which can be seen from Figure \ref{eeuf_ratio} by comparing the dotted and solid curves), this appears to be only a minor effect. In these models, we use disk blackbody photons as seed photons for Comptonization, with their peak temperature, $T_{\rm diskbb}$, as a free parameter.

The inclinations of the disk, the hot corona and the reflection component are all kept at the best-fit value of the binary, $i=36\fdg 4$. In the disk models, the inner radius is at the ISCO at a respective value of the spin.

\section{Results}
\label{results}

\begin{figure*}
\begin{turn}{90}
\begin{minipage}[c][\textwidth][c]{\textheight}
\subfigure{\includegraphics[width=.33\linewidth]{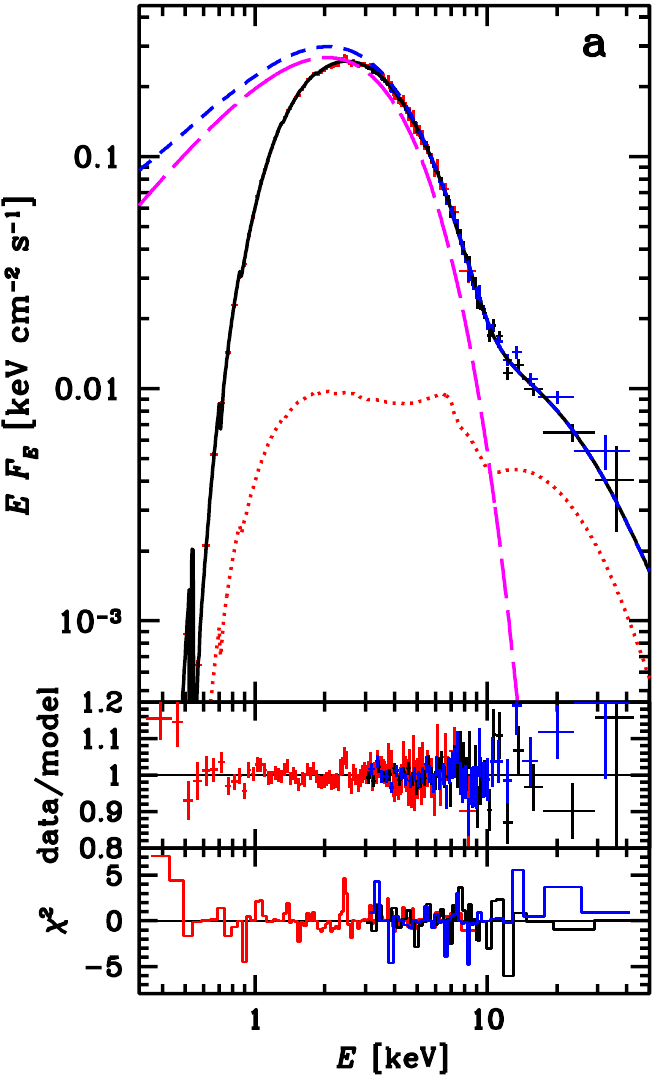}}\hfill
\subfigure{\includegraphics[width=.33\linewidth]{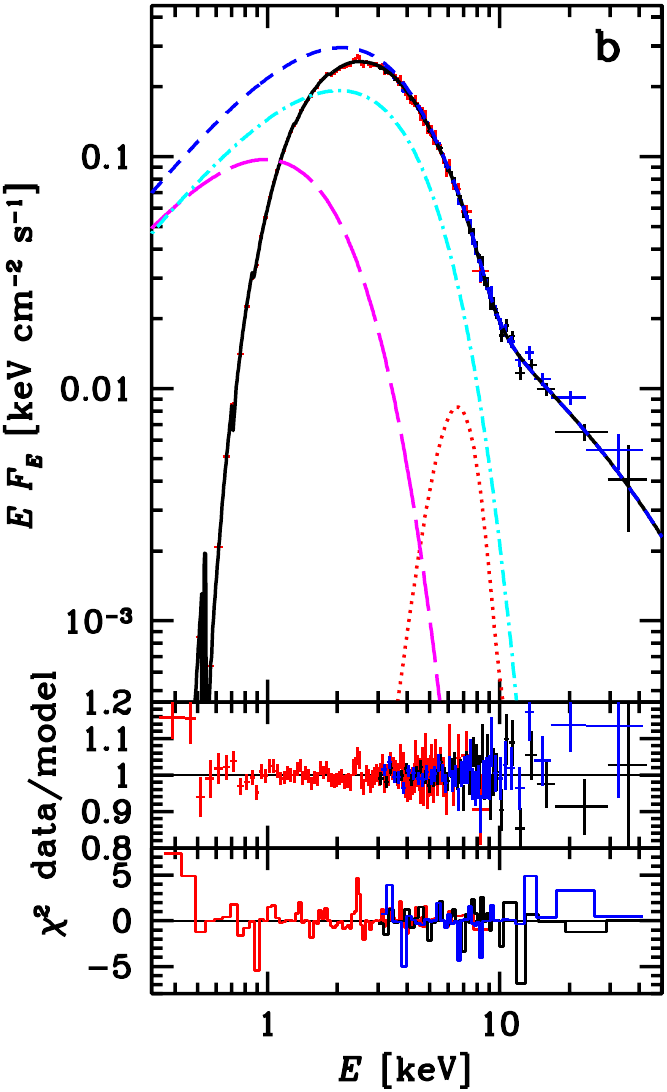}}\hfill
\subfigure{\includegraphics[width=.33\linewidth]{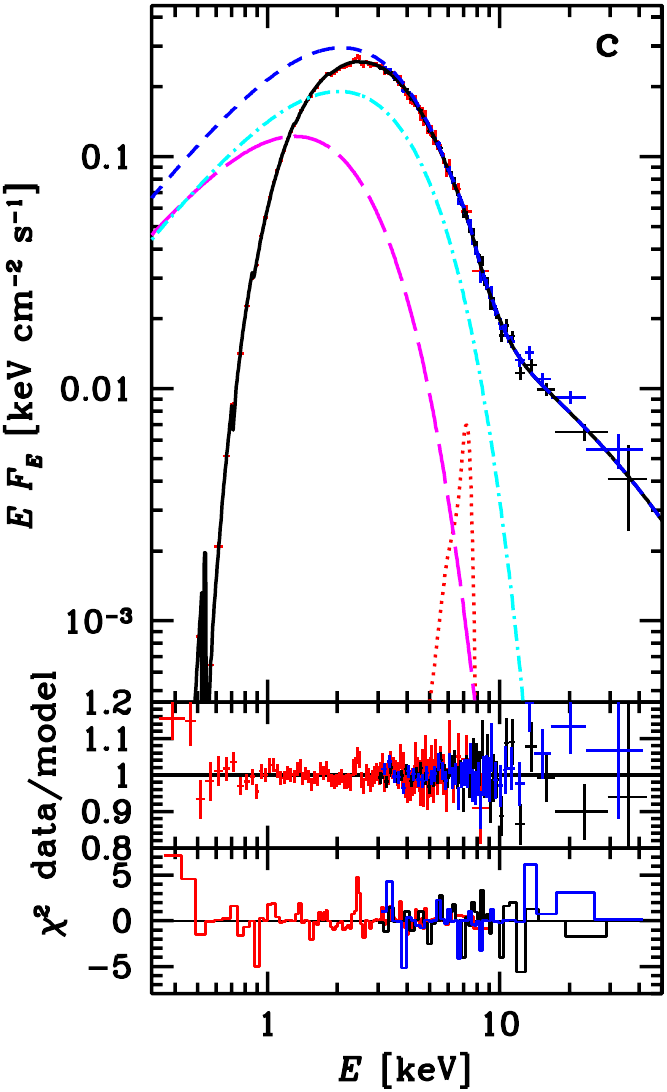}}\hfill
%\subfigure{\includegraphics[width=.33\linewidth]{eeuf2.eps}}\hfill
%\subfigure{\includegraphics[width=.33\linewidth]{eeuf4.eps}}\hfill
%\subfigure{\includegraphics[width=.33\linewidth]{eeuf5.eps}}\hfill
\caption{The \nicer (red) and \nustar (black and blue) unfolded spectra, data-to-model ratios, and $\chi^2$ contributions for (a) Model 2, $a_*=0.94$; (b) Model 4, $a_*=0.40$; (c) model 5, $a_*=0.72$.  The spectra are normalized to \nustar A. The total model spectra are shown by black solid lines, the reflection components are plotted by red dotted curves, and the unabsorbed models are shown by the blue short-dash curves. The magenta long dashes show the disk components before Comptonization, and the cyan dot-dashed curves in (b) and (c) show the disk component after Comptonization in the disk skin.\label{eeuf_ratio}
}
\end{minipage}
\end{turn}
\end{figure*}

Our Model 1 uses {\tt kerrbb2} at $\alpha=0.1$ and it accounts for the reflection/fluorescence feature by using a broad Gaussian line, see Table \ref{fits}. Both the intrinsic and reprocessed emission undergo Comptonization. The fit results are given in Table \ref{fits}. We find the model fits well the data, with the reduced $\chi^2$ of the order of unity. The BH spin is high, $a_*\approx 0.90^{+0.02}_{-0.01}$. We have tested an addition of a narrow Gaussian line at 6.4\,keV, accounting for reprocessing in remote parts of the disk and/or in the wind, but found its addition does not improve the fit ($\Delta\chi^2\approx -0.5$). The unabsorbed bolometric model flux is $\approx 1.06\times 10^{-9}$\,erg\,cm$^{-2}$\,s$^{-1}$, corresponding to the luminosity of $\approx 3.2\times 10^{38}$\,erg\,s$^{-1}$, which is $\approx 0.20\ledd$ (at $X=0.7$) at the best-fit mass of $10.9\msun$. The flux in the unscattered disk emission is $\approx 96\%$ of the total flux. Thus, the state is strongly disk-dominated. 

In our Model 2, we replace the phenomenological Gaussian line by a physical reflection component, {\tt reflkerr}, with the incident spectrum being due to thermal Comptonization with the same parameters as that of the Comptonization of the disk emission, except that we fit the temperature of the seed disk blackbody photons, $kT_{\rm diskbb}$. We assume the irradiation radial profile to correspond to that of the disk dissipation \citep{Page74}. This model gives a somewhat worse fit, see Table \ref{fits}, and Figure \ref{eeuf_ratio}(a) shows the unfolded spectrum including disk and reflection components, the spectrum corrected for absorption, and the data/model residuals. Neither changing the irradiation profile to a power law nor adding a narrow Fe $K\alpha$ line improves the fit. The model gives a very high BH spin, $a_*\approx 0.94^{+0.02}_{-0.04}$. 

One of the parameters of this model is the Fe abundance, which is less than solar in the LMC. Allowing the fractional Fe abundance, $Z_{\rm Fe}$ to be free in the fits, we have found its local minimum at $\approx 1_{-0.4}$, and the global minimum at an unphysical $Z_{\rm Fe}\approx 8$ at $\Delta\chi^2\approx -0.6$. The used rest-frame reflection model assumes a low reflector density, which, for the fitted ionization parameters, leads the irradiating flux to be strongly underestimated, which, in turn, leads to Fe abundance fitted in XRBs to assume unphysical values \citep{Garcia18n}. Therefore, we have kept $Z_{\rm Fe}=1$ in the fits. 

As a variant of Model 2, we replaced {\tt reflkerr} by {\tt reflkerrvnth}, in which the corona is corotating with the disk, and the inner and outer radius and the minimum and maximum heights of the corona are specified. We have found only weak changes in the best-fit parameters with respect to Model 2 for wide ranges of the corona radii and heights, with $\chi^2_\nu\approx 251/265$.

Then, as a variant of Model 1, we replaced {\tt kerrbb2} with {\tt kerrbb}, which is parametrized by the color correction, which we set to the canonical value of $\kappa=1.7$. We obtain very similar parameters as for our Model 1 (which assumed $\alpha=0.1$ and the dissipation in the disk midplane), in particular we find $a_*\approx 0.90^{+0.03}_{-0.01}$, almost identical to the range shown for Model 1 in Table \ref{fits}. This closeness of the obtained parameters implies that the model {\tt kerrbb} with $\kappa=1.7$ is closely equivalent to {\tt kerrbb2} with $\alpha=0.1$.

In Model 3, we consider the effect of increasing the color correction in {\tt kerrbb}, which corresponds to the dissipation being more extended and thus with a stronger role of the Compton scattering within the disk. We set $\kappa=2$, and find that this results in a significant decrease of the spin parameter, to $a_*\approx 0.77^{+0.07}_{-0.13}$. 
 
\begin{figure}[t!]
\centerline{
\includegraphics[width=7cm]{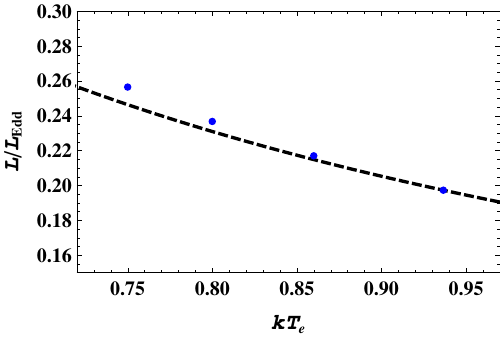}}
  \caption{A theoretical calculation of the dependence of the temperature of the Comptonizing skin, $kT_{\rm e}$, on the luminosity based on the assumption that the Thomson optical depth of the skin, $\tau_{\rm T}$, is proportional to the mass flow rate in the underlying disk, $\dot M_{\rm disk}$. The point for the highest $kT_{\rm e}$ corresponds to our Model 5. We then increase $\dot M_{\rm disk}$ by factors 1.1, 1.2 and 1.3 and calculate the temperature from the requirement of $L\propto \dot M_{\rm disk}$. The resulting dependence can be approximated as $L\propto T_{\rm e}^{-1}$, shown by the dashed curve. It qualitatively reproduces the observed dependence of $L\simprop T_{\rm disk}^{-1}$ (where $T_{\rm disk}$ was compared to the data without including skin Comptonization) for a sample of \xte observations of LMC X-1 shown in fig.\ 2 of \citet{GD04}.
}\label{lum_kte}
\end{figure}

In Models 4 and 5, we consider modifications of the disk emission by a warm skin on top of it, as discussed in Section \ref{model}. The two models differ in the reflection component, using either a Gaussian feature or an intrinsically narrow line broadened relativistically as in {\tt reflkerr}, named {\tt reflkerrline}. The rest-frame energy of the latter is limited to that of the Fe K$\alpha$ line for various ionization levels. As expected, the fitted line energy corresponds to a strongly ionized plasma at $kT\sim 1$\,keV, with the dominant Fe{\sc xxvi} ionization state. The inclusion of the warm Comptonization strongly affects the inferred spin. In Model 4, the spin is in the 0.08--0.81 range. In Model 5, $a_*$ is in the 0.39--0.75 range. Figures \ref{eeuf_ratio}(b, c) show the unfolded spectra and residuals, as well as the intrinsic disk components and those after Comptonization in the disk skin for these two models.

In all of the models, we have found positive residuals at $E\lesssim 0.5$\,keV with the data/model ratio $\approx$1.15, in spite of our fitting the O and Fe abundances (see above). They may be due to some of the assumed elemental abundances in the LMC absorption component being still not fully correct. Alternatively, the shape of the disk component could be modified at low energies due to the re-emission of the irradiating coronal emission, which effect is not included in our models, as discussed in Section \ref{model}. Also, it is possible that the \nicer calibration suffers from some inaccuracies in this range. Furthermore, we see a spike in the residuals between 2 and 3\,keV in all of the models, which appears to be due to an instrumental issue.

\section{Discussion}
\label{discussion}

\subsection{Dissipation in the disk surface layers}

We see that the spin of the BH in LMC X-1 is very model-dependent. Using a GR model for the disk emission (for $\alpha=0.1$) and including Comptonization in a more accurate way than before, we obtain $a_*\gtrsim 0.9$, still in agreement with the previous results. However, this conclusion hinges of the assumption of the standard disk model \citep{SS73, NT73}, with the viscous dissipation localized close to the disk midplane. At the color correction of $\kappa =1.7$, we obtain the same results as for the full disk model with $\alpha=0.1$. But increasing $\kappa$ to a plausible value of 2 (\citealt{Salvesen21}; which may correspond to $\alpha\sim 1$) leads to a decrease the spin to $a_*\approx 0.77^{+0.07}_{-0.13}$.

As discussed in Section \ref{intro}, it is likely that the disk dissipation occurs not only in its midplane but also in surface layers, making them warm and dominated by scattering. We thus consider models with a warm Comptonizing skin on the surface of the standard disk. We consider two such models, differing in the treatment of the reflection feature. When approximating it by a broad Gaussian, we find the spin parameter to be relatively low and only weakly constrained, in the 0.08-0.81 range. This is also the model with the lowest $\chi^2$. When replacing the Gaussian line by an Fe K line narrow in the rest frame, the spin parameter is in the 0.38--0.75 range. Thus, the current best data for this source still do not allow us to determine its BH spin, in spite of the distance to the source being known quite precisely, as well as with the BH mass estimate appearing reliable. 

The optical depth of the skin is $\tau_{\rm T}\approx 24$, 18, in Models 4 and 5, respectively. We note that these values were obtained with {\tt thcomp}, which assumes a spherical geometry with a sinusoidal distribution of the seed photons along the radius \citep{ST80}, while the actual geometry is a slab. For a slab, the equivalent optical depth is a few times lower than that for the above spherical geometry \citep{ST80}. At $L\sim 0.2 L_{\rm Edd}$ and $\alpha\sim 0.1$, the disk Thomson optical depth is $\sim\! 10^3$ \citep{SS73}, as well as it increases when a fraction of the dissipated power is released in a corona \citep{SZ94}. Thus, the optical depth of the skin represents only a tiny fraction of the total optical depth.

Only low upper limits on the X-ray polarization have been found from LMC X-1 \citep{Podgorny23}. As shown in that work, it is consistent with the presence of a standard accretion disk viewed at a modest angle \citep{Cha60}, also with some cancellation of the disk polarization with the polarization of the optically thin corona with the polarization angle perpendicular to the disk one. However, an optically-thick scattering skin has the same polarization properties as the disk \citep{Cha60}. Thus, the presence of warm disk skin is consistent with the polarization results. 

We note that our Models 1 and 4 with the reflection features approximated phenomenologically as broad Gaussian lines give better fits than the corresponding Models 2 and 5, using more physical reflection models. In the former case, the {\tt xillver} model (used for Model 2 in the rest frame) was calculated for low-temperature, high-density media, and it is likely to be quite inaccurate at the high temperatures of the disk in LMC X-1, up to $\sim$1\,keV. In the latter case, using only the dominant Fe K component is likely to be inaccurate as well. Thus, the phenomenological Gaussian components used in Models 1 and 4 are still likely to represent better approximations to the actual features.  

Also, we need to check the energy balance of these solutions. Namely, an isotropically-emitting plasma above a disk radiates a half of its flux toward the disk. Then, a fraction $(1-a)$ of that flux, where $a$ is the scattering albedo, is absorbed and re-emitted, contributing to the disk emission. Thus, self-consistency requires that the power generated in the plasma is not higher than $2/(1-a)$ of the disk emission \citep{HM91}. In Model 4 (which has the weakest disk component and the lowest $\chi^2$) and 5, the unabsorbed fluxes emitted by the disk, the disk + the warm skin, and by the entire source are $\approx$(3.3, 4.2), (6.7, 6.3) and $(10.7,\, 10.6)\times 10^{-10}$\,erg\,cm$^{-2}$\,s$^{-1}$, respectively. Thus, the energy generated in the skin is only $\approx$50\% of that generated within the disk. Similarly, the energy generated in the corona (for which the warm skin is the source of seed photons) is only $\approx$40\% of the total one, and the energy balance satisfies that self-consistency condition regardless of the values of $a$. While the difference in the slab and spherical (assumed in {\tt thcomp}) geometries would change details of the feedback between the disk and the warm skin, the effect of it on the emitted spectrum is expected to be minor.

\subsection{Is LMC X-1 a unique source?}

While most BHXRBs in the soft state approximately follow the blackbody dependence of the luminosity on the maximum disk temperature, $T_{\rm disk}$, LMC X-1 is an exception \citep{GD04}. It shows instead an anticorrelation of the luminosity and temperature, which we found to approximately follow the dependence of $L\propto T_{\rm disk}^{-1}$, see fig.\ 2 of \citet{GD04}. We find that such behavior can be naturally explained as an effect of the warm Comptonization layer on top of a standard disk. 

In our modelling of this effect, we use our fitted {\sc xspec} model. We begin with the values of $\tau_{\rm T}$, $\dot M_{\rm disk}$ and $L$ obtained for our Model 5. We then make plausible assumptions that the Thomson optical depth of that layer, $\tau_{\rm T}$, is $\propto \dot M_{\rm disk}$, and $L\propto \dot M_{\rm disk}$. We next increase $\dot M_{\rm disk}$ in our {\sc xspec} model (without the data) by factors 1.1, 1.2 and 1.3. We then adjust the model skin temperature, $kT_{\rm e}$, to that yielding $L\propto \dot M_{\rm disk}$. The resulting dependence of $L(T_{\rm e}$) is shown in Figure \ref{lum_kte}, where we see a negative dependence, with roughly $L\propto T_{\rm e}^{-1}$, shown by the dashed curve. This result is explained by the luminosity being positively correlated with the Compton parameter at $\tau_{\rm T}\gg 1$, $y\equiv (kT_{\rm e}/m_{\rm e}c^2)\tau_{\rm T}^2$, $L\simprop y$. Then, the quadratic increase of $\tau_{\rm T}^2\propto \dot M_{\rm disk}^2$ has to be compensated by a decrease of $kT_{\rm e}$. For Comptonization with $\tau_{\rm T}\gg 1$, $kT_{\rm e}$ precisely determines the spectral cutoff, and $kT_{\rm disk}$ from the modelling with disk blackbody is $\propto kT_{\rm e}$. Thus, our Models 4 and 5 can explain the $L\propto T_{\rm disk}^{-1}$ anticorrelation in addition to fitting well the spectral data.

While the strong departure from the $L\propto T_{\rm disk}^4$ dependence observed in LMC X-1 appears to be an indication of the presence of Comptonization in a warm corona, it also raises a question of the universality of such an effect. Other sources studied in \citet{GD04} do not show such an effect. One possibility is that this is due to the presence of strong wind from the donor in LMC X-1, but not in low-mass BHXRBs. The wind can then contribute to the accretion flow on a large range of radii (instead of the localized flow through the inner Lagrangian point in the case of the Roche-lobe overflow), which, in turn, can form a flow consisting of two parts, a thin disk and a warm/hot coronal inflow. 

In order to consider this issue, we have also fitted a soft-state spectrum of the high-mass BHXRB Cyg X-1, which results are given in Appendix \ref{cygx1}. We have indeed found the same effect as in LMC X-1. An extremely high BH spin was claimed for this source before, $a_*>0.9985$ at the $3\sigma$ level (\citealt{Miller-Jones21} and references therein). With the standard {\tt kerrbb} model and coronal Comptonization, we have found $a_*= 0.997^{+0.003}_{-0.006}$, confirming that finding. However, almost all of that spin would have to be natal due to the short lifetime of that system, $\approx 4\times 10^6$\,yr \citep{Miller-Jones21}, which is very difficult to explain (e.g., \citealt{Belczynski20}). When adding an optically thick Comptonizing skin, the spin parameter is compatible with being much lower, but it is weakly constrained, $a_*\approx 0.82^{+0.16}_{-0.74}$. Interestingly, the quality of the fit is strongly improved as well, with $\Delta \chi^2\approx -39$ for adding two free parameters. This is compatible with the hypothesis that strong winds cause a formation of a warm skin on top of the accretion disk. Since Cyg X-1 has always a considerable high-energy tail in its soft state, it is difficult to calculate $L_{\rm disk}(T_{\rm disk})$, and we are not aware of such studies.

On the other hand, the high mass BHXRB M33 X-7 has an O7III donor, and accretes via wind, while it appears to exhibit the standard $L$--$T_{\rm disk}$ relation \citep{Liu08}, similar to low-mass BHXRBs \citep{GD04}. Then, the BHXRB LMC X-3 also shows $L\propto T_{\rm disk}^4$ \citep{GD04}, but the mass of its evolved donor is intermediate rather than high, and the accretion appears to be via Roche-lobe overflow \citep{Orosz14}. Furthermore, the presence of a similar warm skin was found in the soft state of the low-mass BHXRB GRS 1915+105 \citep{ZGP01}. Apparently, more studies are needed to find both the ubiquity of the presence of warm disk skins in the soft state and its relation to the $L\propto T_{\rm disk}^4$ correlation.

Warm disk coronae with the temperatures and optical depths similar to those found in the present study have been also widely considered to explain soft X-ray excesses in active galactic nuclei (e.g., \citealt{MBZ98, Petrucci20, Ursini20, Xiang22}). Also, indications for the presence of a warm corona were found in the hard state of Cyg X-1 \citep{Basak17}. Thus, warm coronae may be common in different types of accretion flows onto BHs.

\section{Conclusions}

Our main conclusions are as follows.

We have found the measurements of the BH spin parameter of LMC X-1 to be highly model dependent. In the framework of the standard accretion disk model \citep{SS73, NT73, Davis05, Davis06}, in which the dissipation is localized around the midplane, our measurements confirm the previous results of $a_*\gtrsim 0.9$. In particular, the model including relativistically broadened reflection yields $a_*=0.94^{+0.02}_{-0.04}$. 

However, the standard model is well known to fail to explain a number of astrophysical phenomena, e.g., the stability of the soft state of BHXRBs \citep{GD04}. This is in spite of the theoretical prediction of those disks to be radiation-pressure dominated and thus unstable both viscously and thermally. We thus consider alternative models. When simply setting the color correction to 2 (only slightly higher than the canonical value of 1.7), we find $a_*\approx 0.69$--0.84. 

Then, disks can be partially supported by magnetic pressure (e.g., \citealt{Begelman07, Salvesen16a, Begelman17}), \magenta{as well as} the dissipation occurs also in the surface layers. We model that by assuming the disk is covered by a warm and optically thick surface layer. Such models give low spin values, and our model with the lowest $\chi^2$ yields $a_*\approx 0.40^{+0.41}_{-0.32}$. Those models also agree with the inverse disk temperature-luminosity relation found in this source \citep{GD04}. 

In order to check the ubiquity of the presence of warm coronae in the soft state, we have also fitted a spectrum of Cyg X-1 observed by \suzaku. We have also found that including that effect leads to a weakly constrained spin, compatible with being low. In that case, we have also found that including warm Comptonization leads to a large reduction of the fit $\chi^2$, indicating that model is much more likely, and confirming the previous result of \citet{Kawano17}.

Our results provide a possible explanation of the tension between the low BH spins inferred from analyses of merger events detected in gravitational waves and the prevalence of high spins estimated by spectral fitting of BHXRBs. It is possible that the latter will become significantly lower and in line with the merger data when dissipation in surface layers of accretion disks is taken into account.

\section*{Acknowledgements}
We are grateful to P.-O.\ Petrucci and C. Done for valuable discussions, to C. Done and S. Hagen for providing us with the \suzaku data for Cyg X-1 and discussing their results, to Y. Bhargava for help with those data, and to the referee for important comments. We acknowledge support from the Polish National Science Center under the grant 2019/35/B/ST9/03944. 

\appendix
\section{The spin of Cyg X-1}
\label{cygx1}

\begin{table*}
\caption{The results of spectral fitting for Cyg X-1
}
%\vskip -0.4cm
   \centering\begin{tabular}{lcccc}
\hline
Component & Parameter & Model 1 & Model 2 & Model 3\\
\hline
Absorption & $N_{\rm H}$ $[10^{22}]$\,cm$^{-2}$ &  $0.66^{+0.02}_{-0.01}$ & $0.66^{+0.02}_{-0.01}$ & $0.75^{+0.07}_{-0.03}$\\
\hline
Disk & $a_*$ & $0.997^{+0.003}_{-0.006}$ & $0.986^{+0.005}_{-0.014}$ & $0.82^{+0.16}_{-0.74}$\\
& $\dot M_{\rm disk}\, [10^{17}{\rm g\,s}^{-1}]$ & $2.0^{+0.2}_{-0.4}$ & $1.3^{+0.3}_{-0.1}$ & $1.9^{+4.8}_{-0.5}$\\
& $\kappa$ & 1.7 & 2.0 & 1.7\\
\hline
Coronal  & $y$ & $0.20^{+0.03}_{-0.06}$ & $0.07^{+0.01}_{-0.01}$ & $0.06^{+0.15}_{-0.01}$\\
 Comptonization & $kT_{\rm e}$ [keV] & $32^{+15}_{-11}$ & $76^{+7}_{-6}$ &  $101^{+18}_{-50}$\\
&$f_{\rm cov}$ & $0.11^{+0.10}_{-0.03}$ & $1.00^{+0}_{-0.21}$ & $1.00^{+0}_{-0.94}$\\
\hline 
Disk skin & $\tau_{\rm T}$ & -- & -- & $14.0^{+6.4}_{-12.6}$ \\
Comptonization & $kT_{\rm e}$ [keV] & -- & -- & $0.62^{+5.16}_{-0.10}$ \\
\hline
Cross-calibration 
& $K_{\rm PIN}$ & $0.87^{+0.03}_{-0}$ & $0.87^{+0.03}_{-0}$ & $0.87^{+0.08}_{-0}$\\
\hline
& $\chi_\nu^2$  & 355/285 & 352/285 & 316/283\\
\hline
\end{tabular}
\tablecomments{Models 1 and 2 = {\tt const*tbabs*comppsc(kerrbb)}, Model 3 = {\tt const*tbabs*comppsc(thcomp(kerrbb))}. $i=27\fdg 5$, $M_{\rm BH}=21.2 \msun$, $D= 2.2$\,kpc, and $K_{\rm PIN}\geq 0.87$ are assumed.}
\label{fits_cygx1}
\end{table*}

\begin{figure*}
\centerline{
\includegraphics[width=8.1cm]{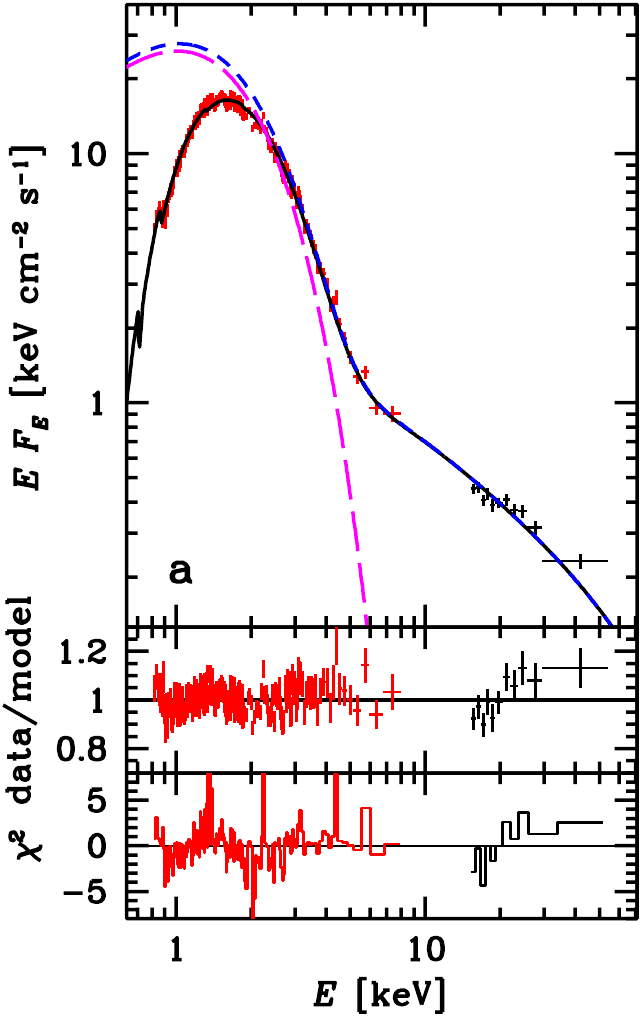}\hfill
\includegraphics[width=8.1cm]{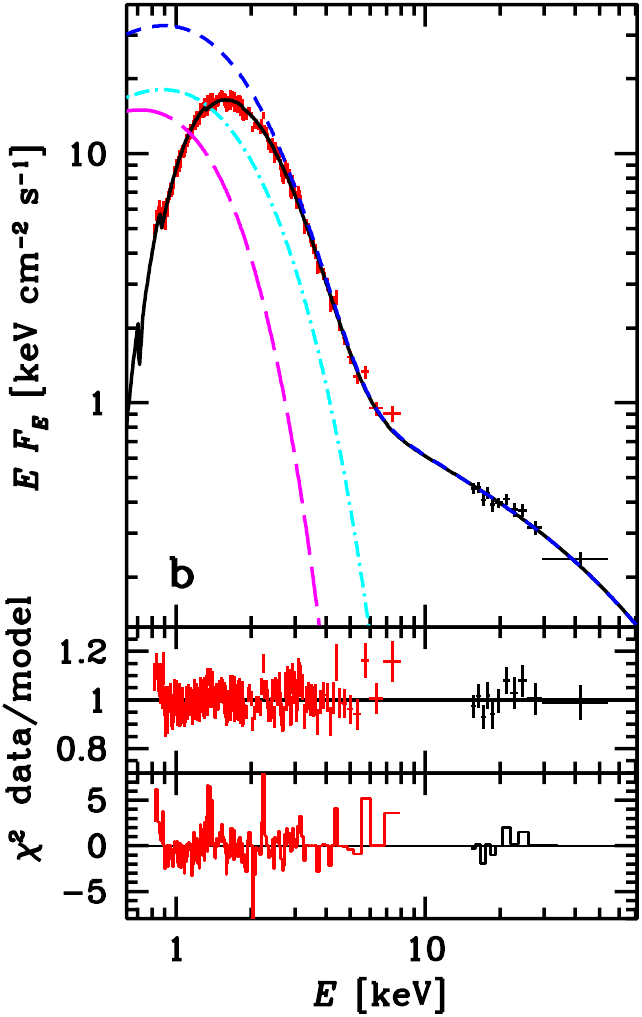}\hfill}
  \caption{The XIS0 (red) and PIN (black) unfolded spectra, data-to-model ratios, and $\chi^2$ contributions for Cyg X-1 fitted (black solid lines) with (a) Model 1, $a_*=0.997$; (b) Model 3, $a_*=0.82$.  The spectra are normalized to XIS0. The unabsorbed models are shown by the blue short-dash curves. The magenta long dashes show the disk components before Comptonization, and the cyan dot-dashed curves in (b) show the disk component after Comptonization in the disk skin.
}\label{eeuf_ratio_cygx1}
\end{figure*}

We address now the question of the applicability of our models to other sources than LMC X-1. We consider as an example the high-mass X-ray binary Cyg X-1. We have accurate determination of its binary parameters, including its binary inclination, $i=27\fdg 5_{-0.6}^{+0.8}$, the mass of the BH, $M_{\rm BH}=21.2\pm 2.2 \msun$ and the distance to the source, $D= 2.22^{+0.18}_{-0.17}$\,kpc (given here as the median values with 68\% uncertainties; \citealt{Miller-Jones21}). Hereafter, we assume the above median values. An extremely high BH spin was claimed for this source before, $a_*>0.9985$ at the $3\sigma$ level (\citealt{Miller-Jones21} and references therein). We study here the \suzaku data used before to measure the BH spin by \citet{Kawano17}, which data show the weakest high-energy tail beyond the disk blackbody observed in the soft state of Cyg X-1 up to 2016 (see \citealt{Kushwaha21} for an update). That spectrum is denoted as B4 by \citet{Kawano17}, and we use it as reduced by those authors. They also considered the effect of thermal Comptonization on the spin value, though using a different and less self-consistent model than that used by us. They found that including it reduces $a_*$ from $0.95\pm 0.01$ to $0.80_{-0.30}^{+0.08}$. Admittedly, this spectrum has a limited statistics and a gap between the XIS and PIN energy coverages; see \citet{Kawano17} for details. We still use that as a test of the applicability of our models to other sources. Analysis of simultaneous \nicer/\nustar data for the soft state of Cyg X-1 will be presented elsewhere (work in preparation). 

We use the {\tt kerrbb} model instead of {\tt kerrbb2} since we found that fits with the latter yield values of $\dot M_{\rm disk}$ outside the tabulated range. As we found in Section \ref{results}, the results of {\tt kerrbb2} with $\alpha=0.1$ are very similar those of {\tt kerrbb} with $\kappa=1.7$. For the fitting, we use the energy ranges of 0.8--8\,keV for the X-ray Image Spectrometer (XIS0), and 15--54\,keV for the  Hard X-Ray Detector PIN. While Cyg X-1 in the soft state shows distinct reflection features \citep{Walton16}, here we have found that, due to the limited statistics and the gap in the energy coverage, addition of either a reflection component or a Gaussian line around 6.4\,keV does not improve the models. These features in the studied observation are indeed weak, as found by \citet{Kawano17}, which is due to the weakness of the high energy tail irradiating the disk. As a test of the robustness of our results, we have added to the models the relativistically smeared reflection component (as in Model 2 for LMC X-1) at some fixed normalization values (which increased the $\chi^2$). We have found that this causes the fitted spin parameter to decrease. Thus, our obtained effect of the spin becoming lower and less constrained with adding the skin Comptonization is not an artefact of not including the reflection in the model. 

We have followed a similar modeling procedure to that for LMC X-1, except that we now do not include reflection. This gives us three models, see Table \ref{fits_cygx1}. Model 1 and 2 have only coronal Comptonization acting on the relativistic disk, for which we assume the color correction factors of $\kappa=1.7$ and 2.0. In Model 3, we also include disk skin Comptonization, and assume $\kappa=1.7$. We have encountered an issue of the value of the relative normalization of the PIN spectrum with respect to that of XIS0. The recommended value from the \suzaku team\footnote{\url{http://www.astro.isas.jaxa.jp/suzaku/doc/suzakumemo/suzakumemo-2008-06.pdf}} is $K_{\rm PIN}=1.158\pm 0.013$. We note, however, that while the PIN exposure for this observation is 3447\,s, that for the XIS0 is 215\,s, i.e., only 6\% of that of the PIN. Thus, the true average flux for the XIS0 could be significantly different from that measured, though no systematic trend is seen in the light curve of XIS0 \citep{Kawano17}. In fact, we have found that the fit $\chi^2$ strongly decreases with the decreasing $K_{\rm PIN}$. In the case of our model 1 at $K_{\rm PIN}=1.145$ (i.e., at the lower limit of the recommended range), we obtain $\chi^2_\nu\approx 382/285$. When allowing for a 25\% uncertainty (which appears quite likely) with respect to 1.158, we obtain $\Delta\chi^2\approx -27$ at the lower limit of $K_{\rm PIN}\approx 0.87$, which is a highly significant reduction. Furthermore, the recommended normalization was found for the Crab spectrum, while our spectrum is much softer. We also note that the usual practice in fitting \suzaku data is to allow a free relative normalization of XIS vs.\ PIN. In the case of the hard spectrum of Cyg X-1, which is significantly harder than that of the Crab, \citet{Parker15} fitted the relative normalization of the PIN to be by 20\% different from the recommended value. We thus impose the constraint of $K_{\rm PIN}\geq 0.87$. 

Our results are shown in Table \ref{fits_cygx1} and Figure \ref{eeuf_ratio_cygx1}. We find some wiggles in the residuals for the XIS0 for all of our models; however, they are almost the same as those shown in \citet{Kawano17}, and appear to be of instrumental origin. Similarly, the spikes seen in the $\chi^2$ plots appear to be of the same origin. In Model 1, we obtain a very similar BH spin to that of \citet{Miller-Jones21}, $a_*=0.997^{+0.003}_{-0.006}$. Increasing the color correction to $\kappa=2.0$ leads to only moderate reduction of $a_*$, to $0.986^{+0.005}_{-0.014}$. On the other hand, adding the skin Comptonization at $\kappa=1.7$ leads to a highly significant reduction of $\chi^2$, $\Delta\chi^2\approx -39$ for adding two free parameters. The F-test gives the probability of this improvement appearing by chance of $7\times 10^{-8}$. The spin parameter becomes only loosely constrained and compatible with almost entire 0--1 range, $a_*\approx 0.82^{+0.16}_{-0.74}$ (at the 90\% confidence level). Thus, Cyg X-1 shows a similar effect as in LMC X-1, except that the reduction of $\chi^2$ due to adding a warm disk skin is much larger in Cyg X-1 than in LMC X-1.

After the calculations reported above were finished, \citet{Belczynski23} reported their fits to the same \suzaku data. Their models are similar to ours, with the main differences being the assumption of a fixed $K_{\rm PIN}=1.15$ and the use of the {\tt simpl} model \citep{Steiner09} instead of {\tt comppsc} for coronal Comptonization. The former is an empirical convolution model in which the input photons are upscattered into a power law distribution, with the only free parameters being the power-law index and the scattering fraction. They also assumed $i=30\degr$ and different abundances for the Galactic absorption \citep{AG89}. Similar to us, they used {\tt thcomp} for skin Comptonization, so their model is {\tt const*tbabs*simpl(thcomp(kerrbb))}. That model yields the spin of $a_*=0.81^{+0.07}_{-0.25}$. A fixed $a_*=0.1$ gives $\Delta\chi^2\approx +5$, and still a reasonable fit (see \citealt{Belczynski23} for details). Thus, the results of \citet{Belczynski23} fully confirm our conclusion that the spin of Cyg X-1 is only weakly constrained.

For comparison, our Model 3 (with {\tt comppsc}) with the same assumptions as their yields $a_*= 0.96^{+0.02}_{-0.15}$. This shows a high sensitivity of the fit results to the used coronal Comptonization model. While {\tt comppsc} is more accurate than {\tt simpl}, it assumes a Maxwellian distribution of the coronal electrons while broad-band data show it is hybrid, i.e., containing a high-energy tail \citep{G99}. Then, it is possible that the latter generic upscattering model captures the Comptonization properties for the data at hand better than {\tt comppsc}.

\bibliographystyle{aasjournal}
\bibliography{../allbib} 

\end{document}